\documentclass[10pt]{article}
\usepackage{amsmath}
\usepackage{epsfig}
\usepackage{graphicx}
\usepackage{amsfonts}
\usepackage{amssymb}

\def\tit#1{\begin{centering}\Large\bf #1 \\[5mm]\end{centering}}
\def\aut#1{\centerline{#1}}
\def\add#1{\begin{centering}\it #1 \\[5mm]\end{centering}}

\begin{document}

\tit{Polarized Helium to Image the Lung}

\aut{Mich\`{e}le Leduc and Pierre Jean Nacher}

\add{leduc@lkb.ens.fr - nacher@lkb.ens.fr\\
Laboratoire Kastler Brossel, ENS, 24 rue Lhomond, F75005 Paris\footnote
{Laboratoire Kastler Brossel is a unit\'e de recherche de l'Ecole
Normale Sup\'erieure et de l'Universit\'e Pierre et Marie Curie,
associ\'ee au CNRS (UMR 8552).}
}

\textbf{Abstract} \textit{The main findings of the european PHIL project
(Polarised Helium to Image the Lung) are reported. State of the art optical
pumping techniques for polarising }$^{3}\mathrm{He}$ \textit{gas are
described. MRI methodological improvements allow dynamical ventilation images
with a good resolution, ultimately limited by gas diffusion. Diffusion imaging
appears as a robust method of lung diagnosis. A discussion of the potential
advantage of low field MRI is presented. Selected PHIL results for emphysema
are given, with the perspectives that this joint work opens up for the future
of respiratory medicine.}

\subsection*{Introduction}

In 1994 first images of Magnetic Resonance Imaging (MRI) were published
showing pulmonary cavities of a dead mouse, inflated with optically polarised
$^{129}$Xe gas \cite{Albert94}. Soon after, research teams in the USA and in
Germany, associating atomic and MRI physicists with radiologists, demonstrated
the possibility to image human lung airways using inhaled $^{3}$He
\cite{McFall96,Kauczor96}.

These gases are chemically inert, non radioactive noble gases, both of nuclear
spin 1/2. They are spin polarised prior to inhalation by optical pumping
methods, which means that a large non equilibrium magnetization is obtained by
orientation of the magnetic moments of the gas nuclei. Then the gas is inhaled
in the lungs and imaged by MRI in conventional scanners tuned to the NMR
frequency of the gas atoms. This method allows visualising the intrapulmonary
airspaces filled with the gas, instead of the tissues examined by conventional
proton MRI. The large polarisation of the gas (up to 80\%) compensates for the
lower density of nuclear spins in the gas.

Images of the gas in the lungs can thus be recorded during a breathhold with a
high resolution. Furthermore, the MRI acquisition sequences can be fast enough
to provide dynamical ventilation data. Mapping physical parameters such as
local relaxation times or apparent diffusion coefficients (ADC) indirectly
provide information on the lung function or may reveal subtle changes in the
microstructure of the airways. It is generally recognized that this new
imaging method has unprecedented potential for morphological and functional
analysis of the lung. It is expected to be a novel addition to the arsenal of
pulmonary diagnostic tests and could enable the development of MRI in the
chest, traditionally hampered by the low number density of protons and the
magnetic field inhomogeneities that exist at the air-tissues boundaries in the lung.

\subsection*{PHIL: a European consortium to study emphysema}

It was decided in 2000 to put forward a European consortium, coordinated by
one of the authors (M.L.), to link the efforts of nine teams in five different
countries. Participants were selected for their pre-existing know-how and
complementary competences in atomic physics (Paris, Mainz and Krakow), MRI
methodology and instrumentation (Orsay and Lyon), radiology and respiratory
medicine (Mainz, Sheffield and Copenhagen) and animal model studies (Madrid
and Lyon).

The PHIL consortium (``Polarised Helium to Image the Lung'', \cite{PHIL})
joined efforts to demonstrate the potential and the validity of the new non
invasive method as a diagnostic and prognostic tool for given lung
pathologies: emphysema and selected Chronical Obstructive Pulmonary Diseases
(COPD), such as bronchitis and bronchiolitis. This choice was motivated by the
frequent occurrence of these diseases and the very high cost of their
treatment for society: 10\% of the population and 25\% of the smokers suffer
from COPD, which is the fourth cause of mortality in Europe. The core of the
project was to perform a clinical trial on a large group of patients with the
$^{3}$He MRI method and with conventional techniques: pulmonary function
tests, High Resolution Computed Tomography (HRCT), Krypton scintigraphy. An
important objective of the PHIL project was to provide new tools for the study
of COPD, aiming at differentiation of various types of diseases, as well as
their detection at an early stage, with expectation that in the long range the
findings of the project could lead to monitoring therapeutic treatment.

\subsection*{Polarising $^{3}$He by metastability exchange optical pumping}

The only candidates as inhaled gaseous tracers are $^{3}$He and $^{129}$Xe,
which are spin 1/2 noble gas atoms (isotopes with higher nuclear spins are
disregarded, because they are too sensitive to nuclear relaxation processes).
$^{3}$He has been chosen for several reasons: It provides much larger NMR
signals, as its magnetic moment is nearly three times larger than that of
$^{129}$Xe. The $^{3}$He atoms can be polarised in larger quantities and to
higher polarisation rates. Helium does not cross the alveolar air/blood
barrier and remains confined within the air spaces. It is absolutely harmless
and induces no side effects, whereas xenon is known as a potent anaesthetic at
high concentrations.

$^{3}$He can be polarised by two established methods which rely on optical
pumping techniques. Optical pumping is a very efficient method to control the
atomic spin state through interaction with a resonant light beam carrying
angular momentum. In the presence of a magnetic guiding field $B$ and of weak
relaxation processes, the net result of the repeated light absorption and
re-emission cycles is a change in the relative populations of the involved
atomic sublevels, i.e. creation of spin orientation. For polarising $^{3}$He,
two routes are possible: spin exchange optical pumping (SEOP) \cite{Walker97},
with indirect transfer of angular momentum from a polarised laser beam to the
$^{3}$He nuclei via alkali metal atoms (usually rubidium), and metastability
exchange optical pumping (MEOP) \cite{Nacher85}, with direct transfer of
angular momentum from resonant laser light to $^{3}$He atoms. Each technique
has its own advantages and limitations. For instance, SEOP directly operates
at high pressure but is a slow process (several hours are required to polarise
helium), while MEOP is a much faster process (it only takes seconds to
polarise 100 cm$^{3}$ of gas at 1 mbar), but only operates at low pressure
($\sim$mbar) and requires further compression for its application to lung MRI.
SEOP has been used for several imaging experiments \cite{Chupp01}, but MEOP
has been preferred for human lung imaging in the PHIL project because of its
faster production rates and higher nuclear polarisations.

The principle of a $^{3}$He optical pumping experiment using MEOP is
illustrated in figure~1. 
\begin{figure}[bh]
\centerline{ \psfig{file=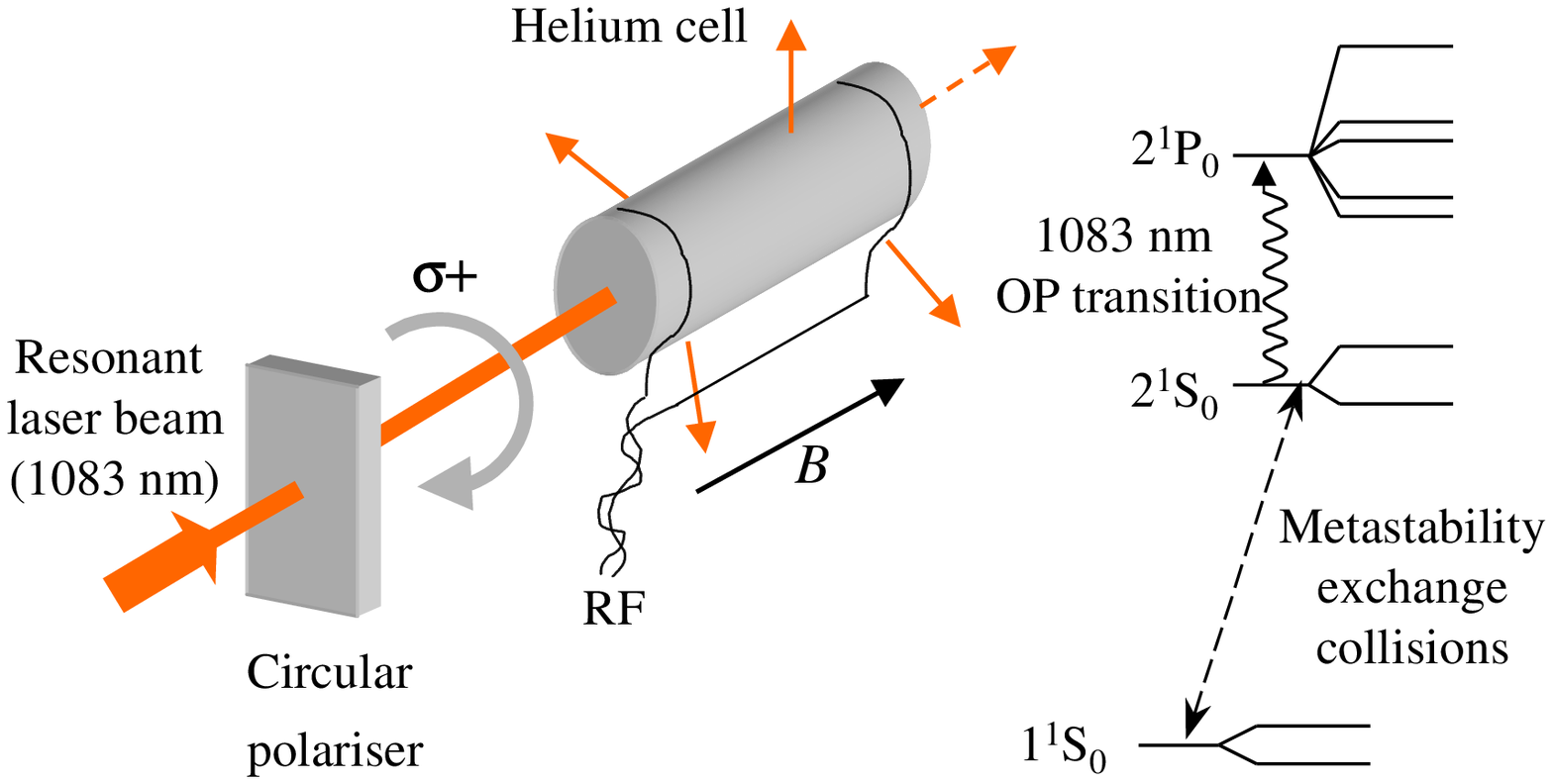, width=10.5 cm, clip= } } 
\caption{Left:
schematic view of an optical pumping setup. In a low-pressure helium-3 gas
cell, a weak RF discharge promotes a small fraction ($\sim$10$^{-6}$) of the
atoms into the excited metastable state 2$^{3}$S, where resonant absorption of
the circularly polarised 1083 nm light can occur. Repeated cycles of
absorption/emission of photons transfer a substantial polarisation in the
2$^{3}$S state. Nuclear polarisation is transferred to the ground state atoms
by metastability exchange collisions. Right: atomic levels of helium-3 and
physical processes involved in the OP cycle.}%
\label{Albert94}%
\end{figure}
A fraction of the helium atoms are excited by a plasma discharge
to a metastable energy state, where they can absorb resonant light to be
optically pumped. For the $^{3}$He isotope in this excited metastable state,
an efficient coupling between the nucleus and the electrons (the hyperfine
interaction) results in a strong entanglement of electronic and nuclear spins.
Therefore optical orientation of the electronic angular momentum
simultaneously induces nuclear orientation as well. This nuclear orientation
is rapidly transferred to the atoms having remained in the ground state
through metastability exchange collisions. This process corresponds to a very
short interaction between an excited atom and a ground state atom, which
results in a fast exchange of the electronic excitations of the colliding
atoms with no change in the nuclear orientations.

In actual gas polarisation devices, a uniform magnetic field (of order 1~mT)
is applied over the whole setup to prevent magnetic relaxation. Powerful fibre
lasers at 1083 nm are used for efficient optical pumping \cite{Tastevin04}.
MEOP-based gas polarisation, which requires compression of the polarised gas,
was developed for PHIL following two routes with complementary advantages:

- Centralised production of very high grade hyperpolarised gas and transport
to the MRI centre. The Mainz group pioneered the development of dedicated
systems for massive production of hyperpolarised $^{3}$He gas \cite{Becker98}.
Sophisticated piston compressors lead to unrivalled performance in terms of
output pressure, gas polarisation and production rate. Such facilities deliver
remarkably large flux (tens of bar litres per day) of high grade polarised
$^{3}$He: over 80\% at the production unit, still reaching 60\% after shipping
by air (from Mainz to Copenhagen or to Sheffield) \cite{vanBeek03}. The
transportation problems were overcome using special containers preventing
magnetic relaxation and appropriate iron-free glass for the storage cells
containing the gas.

- On-site production of gas with compact polarisers implemented next to the
MRI scanner. The delivered gas may be of lower grade ($^{3}$He is diluted in a
neutral gas like nitrogen, or produced with lower polarisation, of order
40\%), but problems with storage and transportation are avoided. The key
element of the system developed by the Paris group is a compact
polarisation-preserving peristaltic compressor that allows production of
polarised gas next to or inside the MRI scanner \cite{Nacher99}. Two different
prototypes of such compact and robust polarisers were built, and are ready for
possible commercial production.

\subsection*{Pulmonary ventilation images: resolution and dynamics}

Figure~2 displays chest MRI images obtained at 1.5 Tesla with conventional
proton MRI (left) and polarised helium MRI (right). 
\begin{figure}[tbh]
\centerline{ \includegraphics[width=11cm]{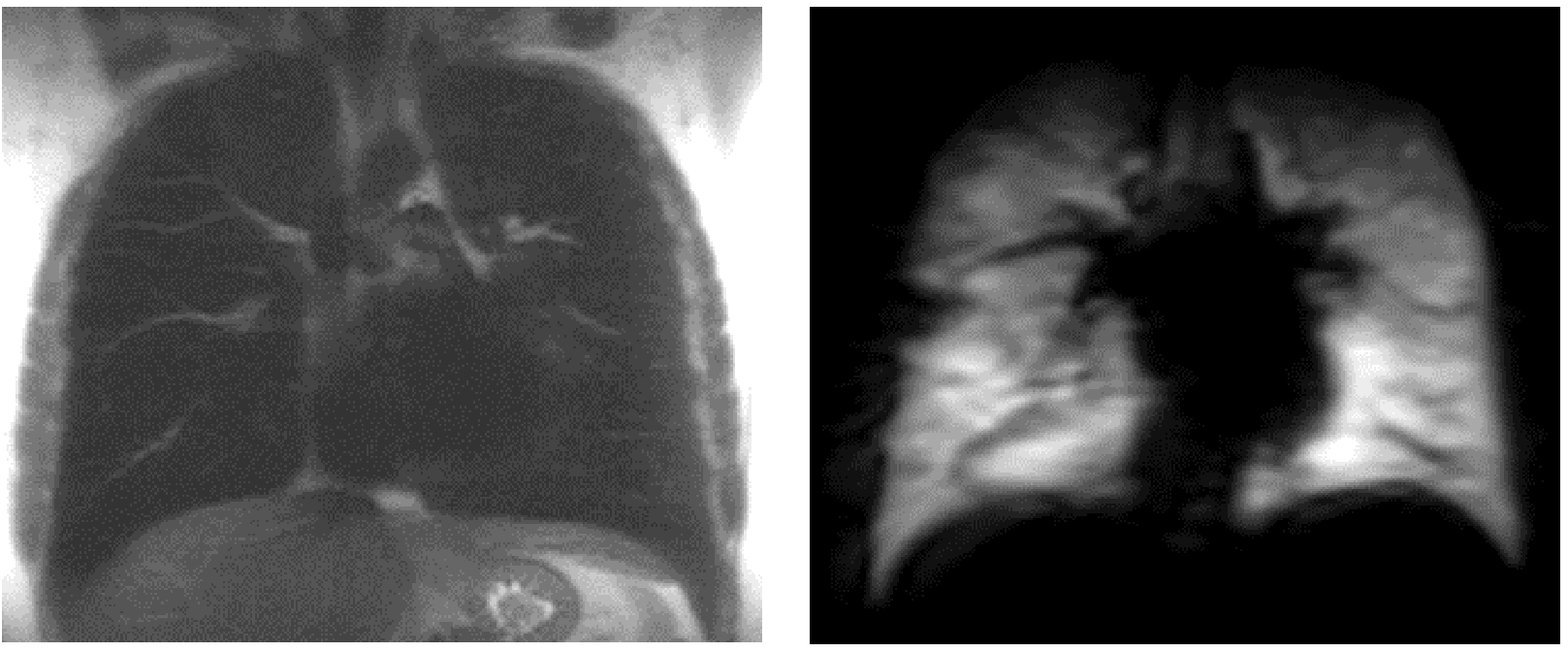} }
\caption{Chest MR
images (1 cm thick slices) of the same normal subject. Left: fast spin-echo
proton image. Right: FLASH $^{3}\mathrm{He}$ image of the gas in the lungs.
The NMR frequency is changed from 64 MHz for protons to 49 MHz for $^{3}$He.
\textit{Courtesy of J.\ Wild, Academic Radiology dept., U. of Sheffield.}}%
\label{fig2}%
\end{figure}
On the proton scan, most tissues are visible, but the lung
parenchyma does not appear: this is due partly to the low density of the lung
tissues, partly to the shortening of the transverse relaxation time $T_{2}$ by
the non-uniform magnetisation of the inhomogeneous parenchyma in the high
magnetic field of the scanner. On the helium scan, the gas-filled pulmonary
air spaces are the only visible part of the chest. Shortening of $T_{2}$ by
susceptibility effects is less severe thanks to the rapid gas diffusion at the
alveolar scale, a well known motional averaging and line narrowing effect. A
set of 15-20 such slices at adjacent positions in the chest allows
re-constructing a 3D map of gas distribution in the lungs.

Another attractive possibility offered by the $^{3}$He-MRI technique is access
to dynamical parameters of the lung ventilation. A typical temporal resolution
of order 0.1~s is required to monitor the respiratory cycle. For ultra-fast
imaging, a radial acquisition strategy combined with a sliding window method
was developed. This method, referred as SPIRO (Sliding Pulmonary Imaging for
Respiratory Overview), allows a regional quantification of dynamical
ventilation parameters using a single breath of polarised gas. Figure~3 shows
two series of helium lung images (here 2D projections) taken during a
respiratory cycle. 
\begin{figure}[tbh]
\centerline{ \includegraphics[width=12cm]{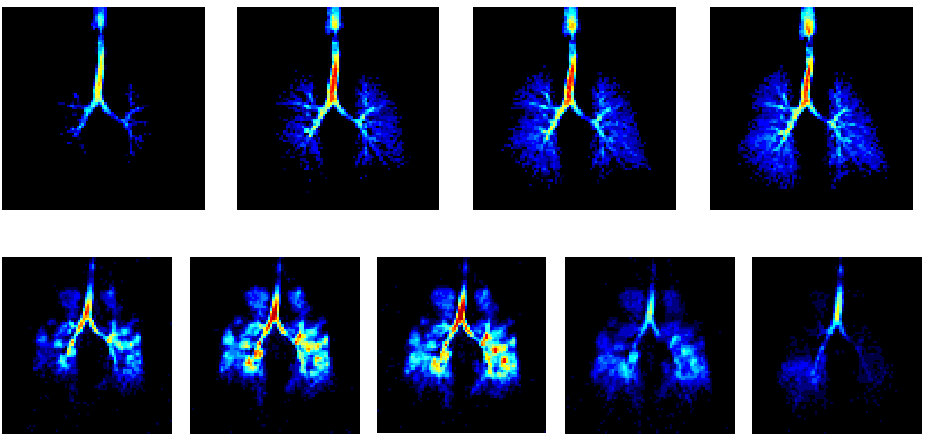} }\caption{Dynamic
images collected using a sliding window radial sequence \cite{Wild03}; the
time interval between displayed images is 0.5~s. Top row: healthy normal
subject, from the first part of an inhalation of 300 ml of $^{3}$He polarized
to 40\%. Bottom row: COPD patient showing regions of ventilation obstruction
in both lungs, particularly in the upper lobes, and a delayed
emptying/depolarization of gas in the lower left lobe which could be
indicative of air trapping. \textit{Courtesy of J.\ Wild, Academic Radiology
dept., U. of Sheffield.}}%
\label{fig3}%
\end{figure}
They show the gas flowing first down the trachea, then into the
bronchi and finally in the peripheral lung. Clear differences are visible
between a normal subject (top row) and a patient suffering from COPD (bottom
row), with wedge-shaped defects indicating that some pulmonary zones are
poorly ventilated. Key functional parameters can be extracted from the images,
such as gas arrival time, filling time constant, and gas volume. Regional maps
of such parameters can be used to assess the degree of the disease severity.

The resolution of these helium images, even if superior to that of competing
techniques, is however intrinsically limited and does not allow visualising
the airways at the scale of the alveoli (of order 0.1 mm). Proton MRI can
reach micrometric resolution for small objects if sufficient signal to noise
ratio is achieved. The same resolution is not possible for helium MRI, due to
the rapid diffusion of the gas inside the airways. MRI relies on frequency
encoding of the position through the use of magnetic field gradients,
sequentially applied along various directions in space. If $G$ is a field
gradient applied along the $x$ direction, the NMR frequency change across a
distance $\delta x$ is $\delta f$ = $\gamma G\delta x$ ($\gamma$=32 MHz/T is
the gyromagnetic ratio of $^{3}\mathrm{He}$). To resolve details of size
$\delta x$, $G$ must be applied for a time $\tau$ such that $\tau\delta f$
$\sim$1, hence $\delta x\sim1/\gamma G\tau.$ When dealing with a gas, the
observation time $\tau$ is actually limited by the diffusion of the gas over
$\delta x$, namely $\tau\sim\delta x^{2}/D$ ($D\sim$80~mm$^{2}$/s is the
diffusion coefficient of $^{3}\mathrm{He}$ in air). Thus the resolution is
limited to $\delta x\sim\left(  D/\gamma G\right)  ^{1/3}$, of order 0.4~mm
for a conventional scanner with typical gradients reaching 0.3~mT/cm.

\subsection*{Apparent Diffusion Coefficient}

In spite of the limitations on image resolution with $^{3}$He-MRI,
measurements of apparent diffusion coefficients (ADC) can provide relevant
information on the lung microstructure. When diffusing freely in air (in the
absence of restricting walls or barriers), helium atoms are displaced on
average by an amount $\sqrt{D\tau}$, e.g. 0.9~mm after a time $\tau$=10~ms. In
the lung, free diffusion occurs at such time scales only in larger airways,
while collisions with the walls of smaller airways and of alveoli reduce
displacements of helium atoms. A time- and scale-dependent ADC is introduced
to characterise this restricted diffusion. It is not easy to relate this
reduced value to a detailed description of the microstructure of the lung at
the alveolar level due to its complexity. However measuring ADC values can be
useful to assess pathological changes. For instance in emphysema, the
restrictions to diffusion are reduced due to expansion of alveoli and
modification of connectivity between them, which results in an increased ADC.

Several NMR\ techniques can be used to measure diffusion coefficients. They
rely on applying a sensitising gradient to imprint a non-uniform magnetisation
pattern during a given time, and on measuring the resulting magnetisation
decay. The ratio of images recorded with and without such a
diffusion-sensitising gradient is used to compute a diffusion map. Figure~4
displays ADC maps obtained in the lung with a standard technique (diffusion
weighting by a bipolar gradient pair). 
\begin{figure}[tbh]
\centerline{ \includegraphics[width=8.8 cm]{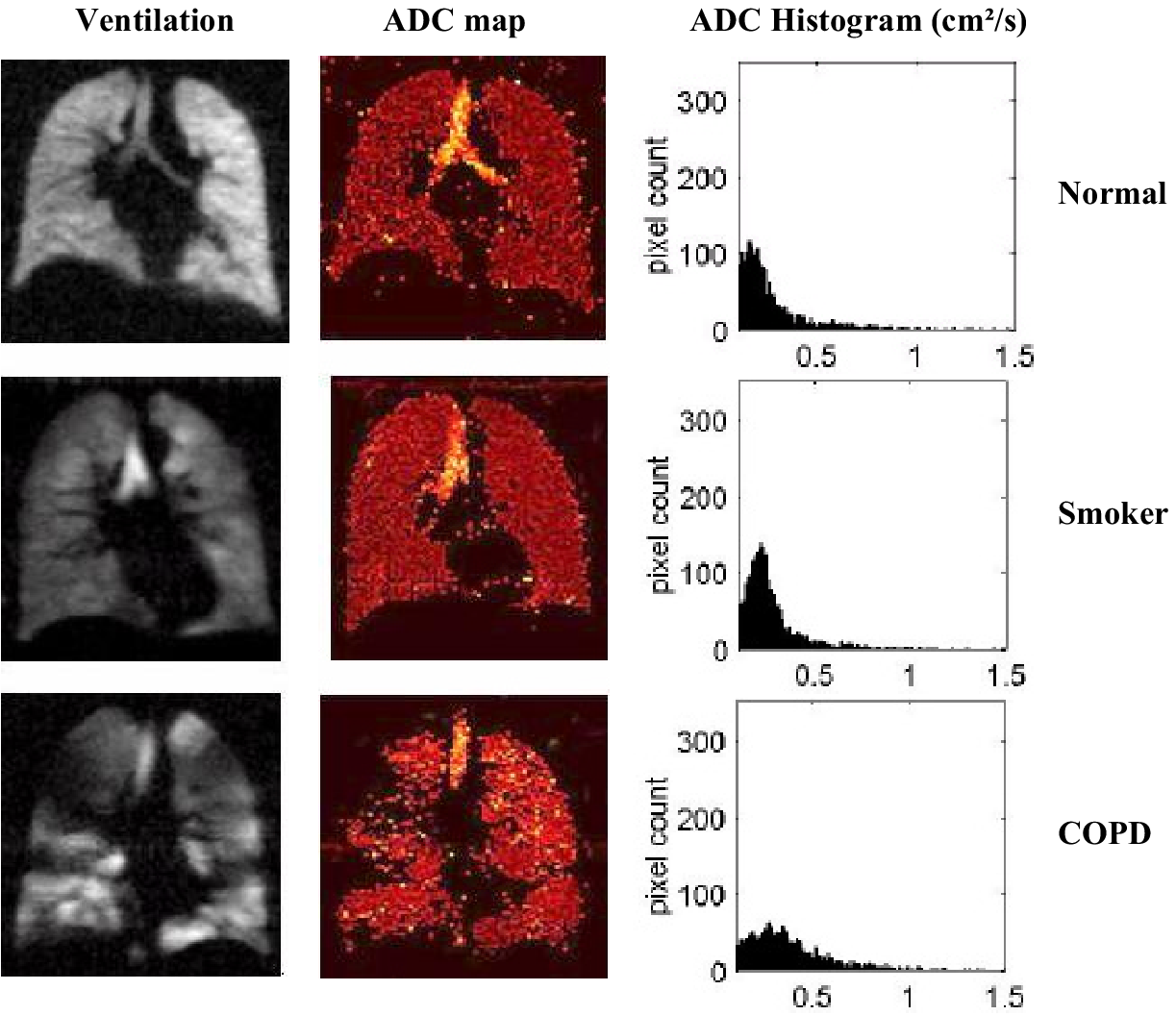} }
\caption{Left to
right: gas distribution images, ADC maps and ADC histograms in 3 subjects (top
to bottom). \textit{Courtesy of J.\ Wild, Academic Radiology dept., U. of
Sheffield.}}%
\label{fig4}%
\end{figure}
One observes an increase by more than a factor two for the mean
ADC value for a COPD patient compared to a normal subject. One also notes a
significant difference between the ADC histograms recorded for an asymptomatic
smoker and a normal subject; this indicates that ADC measurements have a
potential for detection of mild COPD diseases.

\subsection*{Low field versus high field for $^{3}$He MRI?}

In conventional proton MRI, high magnetic fields are used to obtain sufficient
equilibrium proton magnetisation and NMR\ signal. Noise arises from thermal
(Johnson) noise in the patient's tissues and in the NMR receiving coil.
Usually, the former noise source dominates; for a given magnetisation, both
the NMR signal and the noise are proportional to frequency (i.e. to the
field). For $^{3}$He-MRI with optically pumped gas, the magnetisation of the
gas and hence the signal to noise ratio (SNR) do not depend on the field (this
contrasts with the usual proton MRI for which the equilibrium magnetisation,
and hence the SNR, increase linearly with the field). However this breaks down
at very low field (below about 30 mT, depending on the coil size and
technology), where the patient noise decreases below the thermal noise of the coil.

The PHIL Partner in Orsay demonstrated that there is no SNR penalty when using
a low field MRI scanner at 0.1~T \cite{Durand02}. On the contrary, there is an
important benefit of low-field operation: with reduced gradients arising from
the magnetic susceptibility of tissues, transverse relaxation times are much
longer at 0.1~T than at 1.5~T. Weaker field gradients can then be used for
imaging or ADC measurements. This allows measuring diffusion coefficients at
longer time scales, thus providing more information on the connectivity of
lung air spaces. A precise way of measuring ADC coefficients in this case is
through the decay of spin echo trains.

However, shortening of transverse relaxation times can also be viewed as a
useful contrast mechanism since it depends on the lung microstructure. It was
for instance demonstrated to be more sensitive to the degree of lung inflation
at 1.5~T than at 0.1~T \cite{Rochefort04}.

A clear advantage of operating at low field is the reduced cost and increased
flexibility of the imaging system: an open geometry can be designed, and
standing or sitting patients may be examined. Studies at ultra low field in a
home made vertical scanner at 3 mT have started with the PHIL Partner in Paris
\cite{Bidinosti04}. This could open up possibilities of low-cost dedicated
scanners for the screening or follow-up of lung diseases.

\subsection*{Emphysema studies with $^{3}$He MRI}

The PHIL multicentre clinical trial, after upgrading the hardware and software
of the involved MRI scanners according to the $^{3}$He MRI standards, put
forward a standardized protocol to study a significant number of patients and
healthy volunteers for reference. Pre-existing sequences initially used in
Mainz were adapted and optimized at the different sites: ventilation
distribution (breath-hold), apparent diffusion coefficient (breath hold) and
dynamic ventilation (single respiratory cycle). The protocol also included
pulmonary function tests to assess severity scores. Data were recorded in a
data base and finally submitted to a statistical analysis, which provided a
wealth of information, still not fully exploited \cite{Gast04}.

The subjects were divided in three groups: 79 patients with proven COPD, among
whom 17 suffered from alpha-1-ATD (anti-trypsin deficiency) and 37 healthy
volunteers. A score number referring to the degree of abnormality was
attributed to each subject. Comparison between HRCT, MRI findings and
abnormality scores showed that MRI correlates better with pulmonary function
tests than HRCT. The vast majority of MRI findings consisted of wedged shaped
defects, even subjects with normal HRCT showed ventilation defects. The
comparison of ADC with emphysema index and mean lung density was also
exploited. ADC measurements clearly separated the healthy subjects from those
with emphysema of any type. $^{3}$He MRI proved also to be able to distinguish
between different diseases by using a combination of ventilation distribution
and ADC mapping. After assessing all the results, one general finding is that
HRCT is usually more sensitive but MRI is more specific. A positive feature of
the $^{3}$He MRI method is that a single exam provides several unique
complementary informations at the same time (morphology, ventilation defects,
average of air spaces, even in-vivo oxygen concentration and uptake that can
be derived from the local relaxation times).

The animal models studies of PHIL provided very complementary results
\cite{Peces}. Different models of lung diseases were induced in rats with
chemicals such as elastase or cadmium chloride. To some extent these models
mimic human emphysema. The rat groups treated with cadmium showed clear
ventilation defects, corresponding to non-ventilated areas of the partially
collapsed lung. ADC and post-mortem morphometric measurements were compared.
The most significant finding of this work is the excellent correlation between
the ADC data and the morphometric ones, which allow direct quantification of
alveolar size. Complementary to experiments in clinical trials, in which the
available population generally have suffered chronic diseases, emphysema in
animals was mildly developed to evaluate the technique for early diagnosis. In
that perspective, low field scanners could be highly useful, as the system
successfully built in Krakow with a permanent magnet of 0.08~T based on a
permanent magnetic material of new generation. Such low cost scanners are
adapted to imaging small animal lungs and may be especially valuable for
pharmaceutical research.

\subsection*{Conclusion}

In this multidisciplinary work, atomic physicists played a prominent role at
the early stage of the technique. The $^{3}$He polarisation methods were first
developed in the 80's for the needs of polarised quantum fluids physics, for
polarised targets in nuclear physics and more recently for neutron spin
filters at reactors. Finally, in view of application to MRI, the metastability
exchange optical pumping method was pushed to its limits and very appropriate
polarisers were constructed by the PHIL consortium. They bring a satisfactory
solution for clinical implementation of $^{3}$He MRI, either by remote
production of high grade gas or in-situ table-top polarisation. There is still
room for improvement of fast dedicated MRI sequences and signal to noise
enhancement. Great expectation is also related to dedicated low field
scanners, with possible extension of the $^{3}$He MRI lung imaging method to
developing countries.

The PHIL consortium has shown that $^{3}$He MRI can be routinely performed on
a significant number of patients and shows interesting new insights into
common lung diseases, COPD and emphysema. It proves superior in attaining 3D
lung function in the lungs, and is more closely correlated with routine lung
function tests than HRCT. The method has already been tested in other lung
diseases, and more studies are planned or have started. These include lung
transplant assessment, cystic fibrosis (the most common genetic lung disease),
asthma and lung cancer. The technology has proved of huge potential interest
in these areas, because it is non-invasive and does not require ionising
radiation, an extremely important feature, especially in longitudinal
follow-up of children. It could also be very beneficial for the development of
novel therapies, reducing the size of trials and the number of sacrificed animals.

\end{document}